\documentclass[floatfix,onecolumn,pre]{revtex4}

\usepackage{graphicx}
\usepackage{amssymb,amsfonts,amsmath}

\begin{document}

\title{Smectics: Symmetry Breaking, Singularities, and Surfaces} 

\author{Bryan Gin-ge Chen, Gareth P. Alexander, and Randall D. Kamien}
\affiliation{Department of Physics and Astronomy,University of Pennsylvania, Philadelphia, PA 19104-6396, USA}

\begin{abstract}
The homotopy theory of topological defects in ordered media fails to completely characterize systems with broken translational symmetry.  We argue that the problem can be understood in terms of the lack of rotational Goldstone modes in such systems and provide an alternate approach that correctly accounts for the interaction between translations and rotations. Dislocations are associated, as usual, with branch points in a phase field, while disclinations arise as critical points and singularities in the phase field. We introduce a three-dimensional model for two-dimensional smectics that clarifies the topology of disclinations and geometrically captures known results without the need for compatibility conditions.  Our work suggests natural generalizations of the two-dimensional smectic theory to higher dimensions and to crystals. 
\end{abstract}
\maketitle

Many of the features and properties of ordered media, such as crystalline solids, magnets, and liquid crystals, are controlled by their defects; those points or lines of discontinuity at which the order changes abruptly or is ill-defined. For example, defects have played prominent roles in understanding the plastic deformation of solids~\cite{taylor34}, the Kosterlitz-Thouless transition of the two-dimensional XY model~\cite{kosterlitz73}, the dislocation-unbinding melting of solids~\cite{toner81} and the formation of colloidal crystals in nematic liquid crystals~\cite{musevic06}. One of the unifying features of the defects in all of these systems is that they are topological, in the sense that many of their properties depend only on the symmetry of the ordered medium. Based on this concept a general framework has been developed for the identification and classification of the topological defects in any broken symmetry medium in terms of group elements (or more generally conjugacy classes) of homotopy groups of a space of degeneracy, representing the ground state manifold of the system~\cite{klemanmicheltoulouse,mermin,michel,trebin}. This has provided an understanding of the ways in which defects can combine or split, and highlighted the path dependence of such processes when the fundamental group of the ground state manifold is non-Abelian~\cite{poenaru77}. 

It is known, however, that the standard topological theory of defects is incomplete in systems with spontaneously broken translation and rotation invariance, such as crystals~\cite{mermin,trebin,sethna92}. More specifically, a description of such systems that treats disclinations and dislocations (translational defects) and their interactions in a mathematically coherent fashion is missing. A topological classification of treating dislocations alone is available via the use of Burgers' vectors and the like~\cite{chaikinlubensky,sethna}.  The standard way~\cite{klemanmichel,kleman} to add disclinations in these systems is to adjoin extra components representing the orientation of the system to the order parameter. For instance, the order parameter space of the two-dimensional smectic becomes the Klein bottle. But in this formalism, the neat correspondence between the fundamental group of the order parameter space and disclination-type defects breaks down, as we will demonstrate. For instance, it is impossible to have a disclination with index greater than $+1$ in a two-dimensional smectic system~\cite{mermin,poenaru}, yet there is no indication of this in the standard formalism. This breakdown is related to the number of Goldstone modes in these systems, and the restrictions they impose on paths in the ground state manifold. 

The goal of this paper is thus to reconsider the topological theory of defects for systems with broken translational symmetry and argue that ``less is more'', at least for components of the order parameter. We will show that it is mathematically consistent to keep the order parameter as a phase field, of the sort whose topological defects are well-known to generate the Burgers vector for dislocations, and look at disclinations in this picture as defects in the derivatives of the phase field. Modeling disclinations as critical points was considered by Trebin~\cite{trebin}; we complete this approach and present a method which admits dislocations as well by noting that phase fields in layered systems are not maps to $\mathbb{R}$ but rather to periodic, possibly non-orientable spaces. We also suggest connections to singularity theory and Morse theories which can shed further light on our formulation.

We focus on liquid crystalline systems as they have low-dimensional order parameter spaces and use as our prototypical examples the two-dimensional smectic and directed line systems. Smectic liquid crystals consist of rod-shaped molecules that spontaneously form both directional (nematic) order and a one-dimensional density wave, commonly described as a layered system; the spacing between the layers is approximately the rod length, $a$. In two-dimensions, we can picture the layers as a set of nearly-equally-spaced curves lying in the plane. The ground states are characterized by both equal-spacing between these curves and vanishing curvature. The free energy, accordingly, has two terms, the first a compression energy that sets the spacing, the second a bending energy. Because they are periodic, smectics can be expressed as a density wave $\rho({\bf x})$ with wavelength $2\pi/q_0$. Nearly uniform plane waves with wavevector ${\bf q}_0=q_0\hat{\bf q}_0$ can be written as $\rho({\bf x})=\rho_0+\langle\rho_1\rangle \text{Re }e^{i({\bf q}_0\cdot{\bf x}-q_0u({\bf x}))}$, where $u$ is the Eulerian displacement function and $\phi=\hat{{\bf q}}_0\cdot{\bf x}-u$ is the phase field. Drawing the layers at places of maximum density, we see that the layers are those level sets where $\phi$ is an integer multiple of $a$. The energy may be written in terms of the phase field or the displacement, hence we will look at the geometry of the smectic in terms of these functions. We note that smectics have an extra symmetry that the $\phi({\bf x})$ description does not have, namely that $\phi\rightarrow -\phi$ is a symmetry which leads to their lack of orientability. We will discuss this in the following; for now it suffices to say that a configuration of layers arising from $\phi({\bf x})$ carries an orientation. Hence, we call this the directed line system, indeed it is topologically equivalent to wave systems; see, for instance, Nye and Berry's work on wave dislocations~\cite{nyeberry}. Familiar examples of true two-dimensional smectic configurations may be found on your fingerprints. Many other realizations of these systems have been studied experimentally~\cite{harrison,ruppel}; even electronic systems can realize quantum smectic states \cite{kivel}.

Additionally, we introduce a geometrically intuitive model for looking at the topology of smectic defects that allows us to demonstrate our results. Since the phase field is the basic ingredient to study a smectic, it is natural to view the smectic as a surface placed over the plane. The singularities of this surface will then correspond to different defects. Our model motivates a highly geometric expression for the nonlinear smectic free energy functional and allows us to consider compression and curvature energies around defects.

In the following we review the standard topological theory of defects via examples, namely disclinations in nematics and dislocations in smectics. We recapitulate the limitations of the standard formalism when extended to disclinations in the directed line system and the smectic. Next, we present our resolution of these problems in terms of our construct and discuss a geometric formulation of the free energy for two-dimensional smectics and use this to find compression-free defects. Finally, we conclude with applications and a discussion of avenues for extension.

\section*{Defects, Fundamental Groups, and All That}

Systems with broken symmetries are described by non-vanishing order parameters. When the symmetry is continuous there is a degeneracy of ground states continuously connected by the underlying symmetry. Each of these equivalent but different ground states is represented by a different value of the order parameter. The order parameter field is a local measure of the system's order; in other words, each configuration of a medium defines a map from physical space (the plane, in this paper) to the set of order parameters. The goal of the topological theory of defects is to classify the defect structures in physical media by analyzing the properties of these maps which are invariant under continuous deformation, or homotopy.

As an example, recall the familiar XY model of two-dimensional spins on the plane. In the broken symmetry state the spins align along a common direction, making an angle $\theta$ with the $+\hat{x}$ direction. The ground state corresponds to a constant value of $\theta$ and the ground state manifold is thus the unit circle, $S^1$. Locally, the orientation at a point $P$ in any texture corresponds to some point in the ground state manifold and we can think of the ordered state as a map from points in the sample to directions given by points in $S^1$. Upon traversing a closed loop in the sample the orientation must return to its initial value so that $\theta$ can only change by an integer multiple of $2 \pi$, known as the winding number. A non-zero winding indicates that the loop encircles a topological defect, or vortex, whose strength is given by the winding number. Thus, defect states of the two-dimensional XY model are characterized by a single integer. 

These notions are formalized in the homotopy theory of topological defects~\cite{mermin,michel,trebin}. There we start with two groups, $G$ and $H$, which are the symmetry groups of the system in a disordered and ordered state, respectively, so $H\subset G$. The ground state manifold is  the quotient space $G/H$ and the topological defects are classified by the conjugacy classes of the fundamental group, $\pi_1 (G/H)$. In the XY model the disordered state has full rotational symmetry, so $G = SO(2)$, which the ordered state breaks completely, so $H$ is the trivial group. The order parameter manifold is $G/H \cong S^1$, the unit circle, and the fundamental group is $\pi_1 (S^1) = \mathbb{Z}$, reproducing our intuitive result. Two-dimensional nematics are almost identical, except that, since nematics have no heads or tails, we must identify $\theta$ and $\theta + \pi$, so that now $H = \mathbb{Z}_2$. We then have $G/H = \mathbb{R}P^1$ and $\pi_1 (\mathbb{R}P^1) = \frac{1}{2}\mathbb{Z}$, giving the familiar Frank indexing of nematic disclinations~\cite{frank58}. In this way, the topological theory provides a general framework for the classification of defects in any broken symmetry ordered medium. Amongst the key insights it affords is that products of loops in the fundamental group yield information on how defects combine or split and, particularly, the path dependence of these processes when the fundamental group is non-Abelian~\cite{poenaru77}. 

The same approach may be applied, through a ``na\"ive generalization''~\cite{mermin}, to systems with broken translational symmetry, such as directed lines in the plane and two-dimensional smectics. Beginning with the directed line model, the ground state consists of parallel layers, perpendicular to $\hat{x}$ and with equal spacing $a$ and a well-defined orientation as in Figure 1(a). The symmetries of the ground state consist of a translation of the layers by $a$ in the $\hat{x}$ direction, arbitrary translations in the $\hat{y}$ direction, and rotations by $2\pi$ of the system. Thus $G$ is the two-dimensional Euclidean group consisting of two translations and a rotation, $H=\mathbb{R}\times\mathbb{Z}$ and $G/H=S^1\times S^1$, the two-torus, where the first factor is translations {\sl modulo} $a$ and the second is rotations {\sl modulo} $2\pi$. 
Coordinates for the ground state manifold are naturally provided by the phase $\phi$ (acted upon by translations) and the direction $\theta$ of the layer normal (acted upon by rotations). If we consider homotopy classes of loops with non-zero winding only in the $\phi$ direction we recover dislocations. In particular, the usual construction for the Burgers vector via counting layers is equivalent to looking at the winding number of $\phi$, since we draw a new layer for every multiple of $a$ in $\phi$. The fundamental group of the torus is $\mathbb{Z} \times \mathbb{Z}$, so in this generalization, defects in the full system should be characterized by pairs of integers $(m,n)$, $m$ corresponding to dislocation charge and $n$ corresponding to disclination charge. 

In the case of the smectic, the ground state is again a set of parallel lines perpendicular to $\hat{x}$, but without orientation. Therefore, the symmetries include translation by $a$ in the $\hat{x}$ direction, and rotations by $\pi$ owing to the nematic-like symmetry. A rotation by $\pi$ followed by a translation by $\delta$ is equivalent to translation by $-\delta$, which leads to a twist in the order parameter space. Again with $\phi,\theta$ being the coordinates of the ground state manifold, we must now identify $\phi$ with $-\phi$ when $\theta\rightarrow\theta +\pi$. The order parameter space is no longer the two-torus, but is the Klein bottle. To construct the fundamental group of the Klein bottle we start with the free group of two elements $S$, representing a shift by $a$, and $F$, a rotation or flip by $\pi$. Note, however, that $FS^{-1}F^{-1}=S$ and it follows that the fundamental group is $\langle S,F|FSF^{-1}S=e\rangle$, where $e$ is the group identity. This group is the semidirect product $\mathbb{Z}\rtimes\mathbb{Z}_2$~\cite{kleman}, so one might classify loops by two numbers $(m,n)$, where $m$ counts the number of dislocations and $n$ counts the number of disclinations. Algebraically, one can show that if $n$ is odd, $(m,n)$ is conjugate to $(m+2,n)$, {\it i.e.} $S^{-1}(FS^m)S=FS^{m+2}$. This can be interpreted graphically and elegantly as the generation of a dislocation pair in the presence of a $+1/2$ disclination, as in Figure 3 of~\cite{kuriklavrentovich}.

Though the algebraic structure we have presented is the natural generalization of the standard study of topological defects~\cite{klemanmichel,klemanlavrentovich}, it fails to correctly characterize the defects of either system. Mermin recognized this in~\cite{mermin} and pointed out the difficulty of using the standard approach when translational symmetry is broken. Part of the issue is that the homotopy theory predicts entire classes of defects that are not present in the physical system. Po\'enaru proved that directed line and smectic models cannot have disclinations with index greater than $+1$ \cite{poenaru}, and yet they are predicted by the homotopy theory. More generally, the problem lies in the allowed changes of the order parameter as one moves along a path in the sample. For a system with liquid-like order, such as the nematic, any variation in the order parameter can be accommodated, {\sl i.e.} any path in the ground state manifold can be mapped onto the specified path in the sample. However, the same is not true for translationally ordered media, like the smectic. A continuous symmetry implies that two degenerate ground states can be connected via an arbitrarily low energy Goldstone excitation.  This implies that the dimensionality of the allowed excitations corresponds to the number of Goldstone modes.  Although the smectic breaks both rotational and translational symmetry, it has only one Goldstone mode, the Eulerian displacement field $u({\bf x})$, a signal of something different. Not any path in the ground state manifold can be realized, but only those which correspond to this mode. The reduction in the number of Goldstone modes and consequent restriction of the realizable paths in systems of broken translational symmetry arises because the rotational and translational degrees of freedom are coupled into a composite object~\cite{toner81,sethna92,lowmanohar}, in a manner akin to the Higgs mechanism for gauge fields~\cite{gauge_comment}. Only when the number of Goldstone modes is equal to the dimensionality of the ground state manifold do we have sufficient freedom to be able to match any path in the ground state manifold to any path in the sample, as is required for the application of the standard homotopy theory. 

Our main result is to present a method by which the topological defects of translationally ordered media can be faithfully captured, without the need to remove entire classes of non-realizable defects. The coupling of rotational and translational broken symmetries tells us that we should no longer consider the coordinates $\phi, \theta$ of a point in the ground state manifold as freely variable. Recall that they are both given by the order parameter $\phi({\bf x})$, the former as the value of the phase field at a point and the latter as the direction of its gradient. In order to identify the position in the ground state manifold of a configuration at any point $P$, therefore, we must consider the Taylor series for the configuration $\phi(P+{\bf x})\approx \phi(P)+\left.\nabla\phi\right|_{{\bf x}=P}\cdot{\bf x}+\cdots$~\cite{comment0}. The ground states of the directed line and smectic systems are linear functions $\phi_{\phi_0,{\bf k}}({\bf x})=\phi_0+{\bf k}\cdot{\bf x}$, where $\bf k$ is a unit vector. We assign the $\phi$ component of the local ground state to be $\phi_0$, as usual, and use the directionality of $\left.\nabla\phi\right|_{{\bf x}=P}$ to set the second component, ${\bf k}$. Note how this generalizes the usual procedure. If the ground state has structure up to order $n$, strictly speaking, we should keep data at every point of the configuration also up to order $n$~\cite{comment}. There is an immediate implication regarding the smoothness of the phase field as well. When $\phi$ has a nontrivial winding about a point, we know it must be discontinuous there. Suppose though that a loop has nontrivial winding in $\nabla\phi/\vert\nabla\phi\vert$. In this case the unit normal must be discontinuous, but this only implies that $\nabla\phi$ is discontinuous or zero. It follows that at a disclination $\phi$ is either singular or critical. 

Of course, since any configuration can still be identified locally with a point in the ground state manifold, the defects that do occur will still correspond to non-trivial loops in $G/H$ and we can continue to label them by the winding numbers of these loops. However, the possible defects that can arise should be determined from the homotopy of $\phi({\bf x})$ and $\nabla\phi({\bf x})$ together, and not from the structure of the ground state manifold. With this construction in hand, we move on to our surface model.

\section*{Surface model}

In this section we provide the first steps towards a theory where the objects being homotoped are not simply paths in the ground state manifold, but rather smooth maps from the physical space to the space of the translational phase field. Thus the theory of defects for these systems should be given in the language of singularity theory~\cite{golubitsky} and flavors of Morse theory~\cite{milnor} which may combine those on circle-valued maps~\cite{farber} and on singular spaces~\cite{goresky}.   Rather than attempt to lay out precise mathematical conditions and proceed into the realm of rarefied mathematics, we discuss our ideas in an intuitive way, exploiting the following fact about smectics in two dimensions. Since the smectic is represented by level sets of a phase field on $\mathbb{R}^2$, we may visualize the phase field as a graph of a two-surface. From this perspective it becomes possible to sidestep the difficulties that we have discussed and to understand, for instance, Po\'enaru's rigorous result afresh. We begin with the simpler case of directed lines.

Starting with a ground state for the lines, $\phi=\phi_0+{\bf k}\cdot{\bf x}$ is the equation for a plane and defines a line for fixed values of $\phi$. Taking level cuts at $\phi=na$, $n\in\mathbb{Z}$ we find lines in the $xy$-plane, with uniform spacing, $a$. Note, however, that the $\phi\rightarrow\phi+a$ symmetry implied by the phase-field construction of the layers means that instead of one plane, we have an infinite stack of equivalent planes each one shifted along the $\phi$ direction by a multiple of $a$ as shown in Figure 1(a). Because of this, we can instead take a cut through all the planes at $\phi=0$, for instance, and equivalently generate the ground state viewing, $\phi$ as an element of $S^1$. As we shall see, this multiplane description is natural from the point of view of defects; they connect the different sheets together to form the required topology. For instance, when taking a Burgers circuit around a $+2$ dislocation, we know that $\phi$ changes by $2a$. This means that the sheets must be connected together with the topology of a helicoid so that dislocations may be thought of as branch points as shown in Figure 1(b). Multiple dislocations are given by surfaces connected by multiple helicoids, {\sl e.g.} Figure 1(c). This approach has been usefully employed to visualize dislocations in waves~\cite{berry,comment2} and we extend it to incorporate disclinations. Disclinations, as we have discussed, are either zeroes or singularities of $\nabla\phi$ and so in this height model, the disclinations correspond to critical points or cusps in the surface \cite{trebin}.  For instance, a $+1$ disclination is represented by a ``mountain'' or ``trough'' in the height function and the disclination sits at the peak or nadir. Concretely, an equally spaced, $+1$ disclination is represented by a regular cone $\phi=\vert{\bf x}\vert$, the level sets at $\phi=na$ giving circular layers with equal spacing, $a$.   Similarly, negative charge disclinations correspond to saddles for creatures with varying numbers of appendages.

The notion of topological equivalence in this model is intuitive. First, note that deformations of these surfaces (with the restriction that horizontally or vertically tangent surfaces are singular) correspond to the class of allowed paths in the $\phi,\nabla\phi$ space that we discussed in the previous section~\cite{fnote}. Hence, results from Morse theory and singularity theory on the topology of these surfaces and their defects up to topological equivalence classes may be regarded as global consequences of the local restrictions developed earlier. As an example, consider Po\'enaru's result on the lack of disclinations of charge higher than $+1$.  In our approach, Po\'enaru's result becomes transparent: two $+1$ disclinations cannot join because two mountains cannot come together on a surface without a mountain pass between them. A mountain pass has a critical point with the geometry of a saddle, and the contours of constant height draw out a $-1$ disclination in the two-dimensional line system. Similar arguments hold for two troughs and even for a trough and a mountain.

Smectics can also have $1/2$ disclinations because they are non-orientable. Thus, in addition to the infinite set of planes normal to $(-{\bf k},1)$ in the line system, the smectic ground state also contains the infinite set of planes normal to $({\bf k},1)$, as shown in Figure 2(a), since these generate the same set of unoriented layers. These two sets of planes intersect whenever $\phi$ is a half-integer multiple of $a$, which reflects the fixed points of the two symmetry operations $\phi\rightarrow\phi+a$ and $\phi\rightarrow-\phi$. At these heights it is possible to cross continuously from one set of planes to the other, thereby changing the local orientation of the layers. Defects again connect the different planes together, although now this can be done in two different ways; either by connecting planes with the same orientation, as in the directed line system, or planes with the opposite orientation. The latter leads to a global unorientability of the smectic and occurs when there are odd half-integer index disclinations. The two prototypical half-integer disclinations are depicted in Figure 2(b) ($+1/2$) and Figure 2(c) ($-1/2$). The $+1/2$ disclination consists of a pair of oppositely oriented planes terminated by a half-cone connecting them, while the $-1/2$ disclination is a three-way junction of pairs of oppositely oriented planes smoothly joined together. These examples provide a visual demonstration of the fact that odd half-integer index disclinations must have their singular points at $\phi = 0$ or $\phi = a/2$. Since these singular points terminate layers and connect the two sets of parallel planes, it follows that the layers must arise as projections from the fixed point values of $\phi$ -- in other words, the presence of charge $1/2$ disclinations forces the defect cores to sit at $\phi=na/2$, $n\in\mathbb{Z}$, or the heights of the level sets and halfway between them.  Properly, for smectics $\phi$ is an element of the orbifold $S^1/\mathbb{Z}_2$, just as it was an element of $S^1$ for directed lines.

The way in which these singularities can join depends on the geometry of the height representation. For instance, consider the dislocation and the {\sl pincement}.  We are used to the idea that a disclination dipole creates a dislocation~\cite{HN}. Indeed in Figure 3(a) we depict a $\pm 1/2$ disclination pair which form a charge $+1$ dislocation. However, consider the {\sl pincement} shown in Figure 3(b). In this configuration the two singular points both lie at the {\sl same} height and hence they can be cancelled. The smectic dislocation cannot be cancelled because the singular layers do not line up. Though this follows from the algebraic structure of paths on the Klein bottle, ({\sl i.e.} $[S^{1/2}FS^{-1/2}]F^{-1} = S$, but $FF^{-1} = e$), the height construction provides a geometric way to see the difference between these two complexions. We are also able to construct a dislocation directly. Since a Burgers circuit around a dislocation changes $\phi$ by a multiple of $a$, it follows that the height function must have the topology of a helicoid. As shown in Figures 3(c) and 3(d) the composite dislocation has the structure of two intertwined half-helicoids. Note that the charge of a dislocation depends on both the sign of the helicoid (left- or right-handed) and the tilt of the layers. A right-handed helicoid moved smoothly around a $+1/2$ disclination remains right-handed, but because the tilt of the layers has reversed the charge of the dislocation changes sign. This is the move that generates the ambiguity {\sl modulo} 2 for the total dislocation charge in the presence of a $1/2$ index disclination~\cite{kuriklavrentovich}. 

In order for dislocation/anti-dislocation pairs to form, a neck connecting adjacent surfaces must be formed. As long as this is smaller than the core size of the defects, it can be created without upsetting the overall topology. As the inchoate dislocation pair separates, these necks expand and, in a manner similar to continuous deformations of Riemann's minimal surface, break up into two oppositely handed helicoids~\cite{riemannref} as depicted in Figure 1(c). Note that this type of process which necessarily includes local surgery for combining defects is implied in the theory of topological defects as well~\cite{mermin}. In smectic systems, however, we must perform local surgery whenever we pull a defect through a layer -- this follows from the above result that the defect cores must remain at a constant height.

\section*{Energetics from a Geometric Point of View}

The height function approach provides a purely geometric way to formulate the nonlinear energetics of a two-dimensional smectic. We start with the compression energy. In terms of the phase field the constraint of equal spacing is embodied in the condition that $\vert\nabla\phi\vert=1$~\cite{SVKN}. Thus any nonlinear strain $u_{zz}$ must vanish when $\vert\nabla\phi\vert=1$. We are free, however, to choose any rotationally invariant function with this property; theories based on different nonlinear strains will differ from each other in anharmonic powers of the strain and only the coefficient of the harmonic term is definition-independent. The height function defines a surface with a unit normal ${\bf N} = (-\partial_x\phi,-\partial_y\phi,1)/\sqrt{1+\vert\nabla\phi\vert^2}$. As we have seen, uniform spacing in the $xy$-plane requires that ${\bf N}\cdot \hat z = \cos(\pi/4)$, where we plot $\phi$ in the $\hat z$ direction. An allowed strain is thus $u_{zz}=( {\bf N}\cdot \hat z - 1/\sqrt{2})$. We note, however that for such a surface, the area measure, $\sqrt{g} = 1/({\bf N}\cdot \hat z)$. Thus we, may generalize the equal spacing condition to the requirement that $g=2$ and choose, for instance, $u_{zz}=g-2$. Indeed, in terms of the height function $\phi$, $g-2=\vert\nabla\phi\vert^2-1$.  

Goldstone's theorem requires that the bending energy, on the other hand, reduce to $\nabla^2\phi$ to harmonic order in $\phi$. Possible nonlinear generalizations include the curvature of the smectic lines $\kappa=\nabla\cdot(\nabla\phi/\vert\nabla\phi\vert)$. Note however, that the mean curvature of the surface defined by $\phi$ is $H=\frac{1}{2}\partial_iN^i$
which reduces to the required expression to linear order in $\phi$. We thus propose the following geometric free energy for the two-dimensional smectic:
\begin{equation}
F=\frac{1}{2}\int dA\left[ B(g-2)^2 + K_1 H^2\right] 
\end{equation}
where $B$ is the bulk modulus and $K_1$ is bending elastic constant.  The free energy is defined as an integral over the height surface that we have introduced and is the Willmore functional in a field~\cite{willmore}. The singularities associated with disclinations and dislocations specify the boundary conditions and topology for the associated variational problem.  

Independent of the energy functional, we know that any equally-spaced two-dimensional smectic has a corresponding height surface with constant ${\bf N}\cdot \hat z$. It follows that the Gauss map of such a surface sweeps out a latitude on the unit sphere and so the Gaussian curvature $K$ of the height surface must vanish. When $K=0$ the surface must be locally isometric to the plane; the surface is either a cone, half of a cone attached to planes, planes, or the tangent developable surface generated by a cylindrical helix~\cite{cac}. Respectively, these four structures have disclination charge $+1$, $+1/2$, $0$, and $+1$. Thus we provide a direct demonstration that it is only possible to have equally spaced layers for these three defect charges.

\section*{Conclusions}

We have outlined an approach to the study of topological defects in systems with broken translational invariance; topological equivalence requires more than just homotopic paths.  One should consider homotopy classes of smooth(er) maps.  Whereas the usual homotopy theory of defects would use continuous homotopy in the $\phi$ and $\theta$ components of the order parameter, we have argued that it is necessary to look at homotopy of $\phi({\bf x})$ and $\nabla\phi({\bf x})$ together.  
Our method focuses on singularities and critical points in a phase field viewed as a height function over $\mathbb{R}^2$. Though the fundamental group of the ground state manifold can be constructed, it is known that when the loops involve both rotations and translations there are homotopy classes of loops implied that cannot be realized from complexions of the physical system. This arises from a mismatch of the dimensionality of the manifold of ground states with the number of Goldstone modes.  To remedy this, we have constructed a local map from the configurations to the ground state manifold by employing Taylor series data at each point. We have shown that the homotopy theory of defects only works for singularities in $\phi$ (dislocations), and not consistently for singularities in $\nabla\phi$. Though the critical points in our theory behave somewhat like disclinations, they are not -- as we demonstrated two $+1$ disclinations cannot be brought together. 
In the case of directed lines, circle-valued maps of the sort we have described above are equivalent~\cite{farber} to closed forms~\cite{xing}. For unoriented lines, that line of formalism may best be extended by considering quadratic differentials instead~\cite{strebel}. Combining these results with ours may be a way of extending our work to the topology of such patterns on surfaces of nontrivial topology (rather than just the plane).  The layer normal of smectics on curved surfaces with planar topology can develop caustics \cite{SVKN}; extending our approach here to those systems is complicated by this additional holonomy and how it interacts with non-orientability near $1/2$ disclinations.

Though we have focused here on two-dimensional smectics and directed lines, our approach easily generalizes to three-dimensional smectics which can be viewed as a height function over $\mathbb{R}^3$ into $S^1/\mathbb{Z}_2$. From this perspective the topological equivalence of edge and screw dislocations becomes particularly vivid; within the four-dimensional space spanned by $(x,y,z,\phi)$ the former are helicoids in $x,y,\phi$ for a fixed $z$, while the latter are helicoids in $x,y,z$ at constant $\phi$. Likewise, the theory of defects in solids can be formulated as maps from $\mathbb{R}^3$ to $T^3/\mathbb{X}$ where $\mathbb{X}$ is one of the 230 three-dimensional space groups. The dislocations will continue to correspond to cycles on $T^3$ while the disclinations will be controlled by the critical points of these maps. Whether the study of singularities can lead to bounds on the total free energy of smectic complexions is an open question as is the general set of rules for the combination of dislocations and disclinations.

\begin{acknowledgments}
We acknowledge stimulating discussions with Jim Halverson, Elisabetta Matsumoto, Carl Modes, and Ari Turner. We are grateful to Koenraad Schalm for bringing reference~\cite{lowmanohar} to our attention. This work was supported in part by NSF Grants DMR05-47230 and DMR05-20020 and gifts from L.J. Bernstein and H.H. Coburn.
\end{acknowledgments}

\begin{figure}
\centerline{\includegraphics[width=15cm]{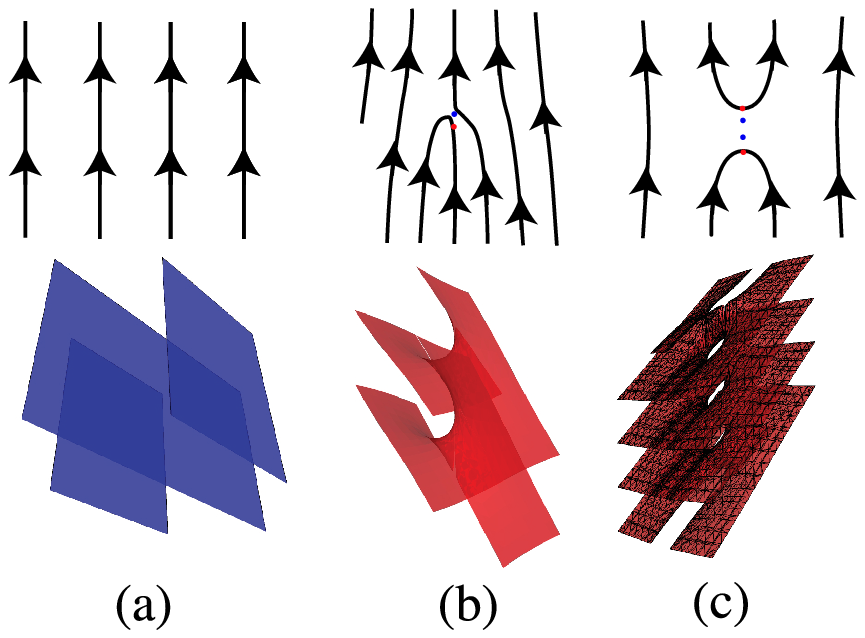}}
\caption{Directed lines in the plane. The top row shows the layers derived from slicing the surface shown in the bottom row at integer values of the height, {\sl i.e.} taking level sets. Red dots indicate the positions of branch points and blue dots indicate the positions of -1 index critical points (saddle point). (a) ground state (b) dislocation / helicoid, (c) two dislocations, note the cylindrical ``hole'' puncturing the planes, reminiscent of Riemann's minimal surface.}
\label{directedfigs}
\end{figure}

\begin{figure}
\centerline{\includegraphics[width=15cm]{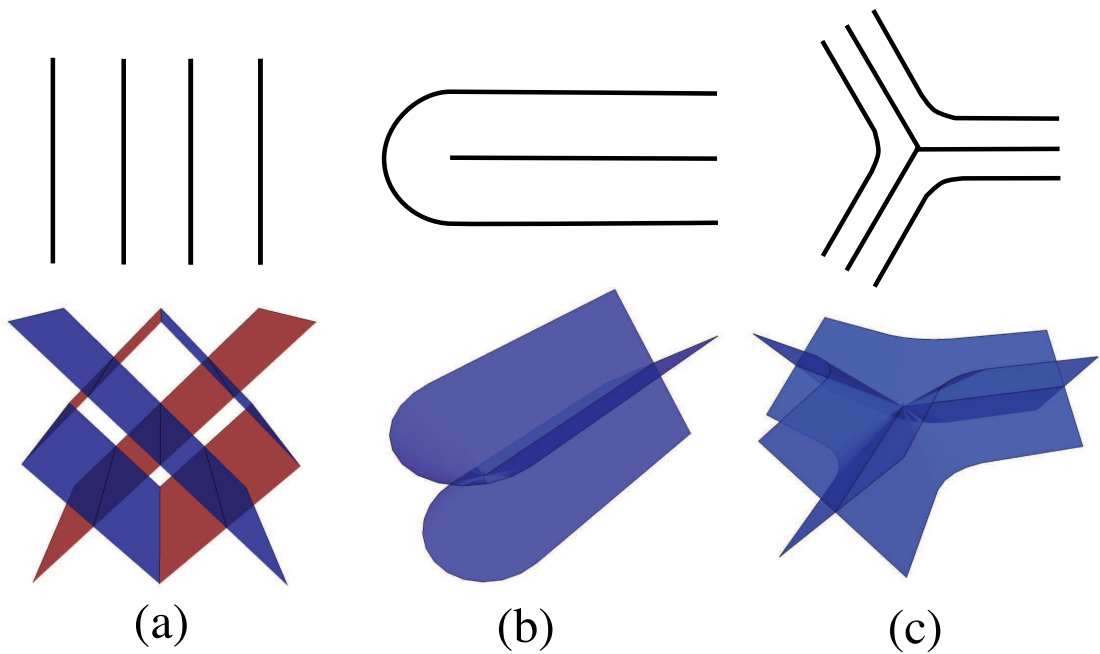}}
\caption{Examples of two-dimensional smectic configurations. The top row shows the layers derived from slicing the surface shown in the bottom row at integer values of the height, i.e. taking level sets. (a) ground state. Note the two sets of planes arising from the $\phi\rightarrow -\phi$ symmetry. (b) $+1/2$ index disclination. (c) $-1/2$ index disclination. These two unoriented singularities join the two sets of planes, so that one can no longer orient the layers as in Figure \ref{directedfigs}.}
\label{smecticfigs}
\end{figure}

\begin{figure*}
\centerline{\includegraphics[width=15cm]{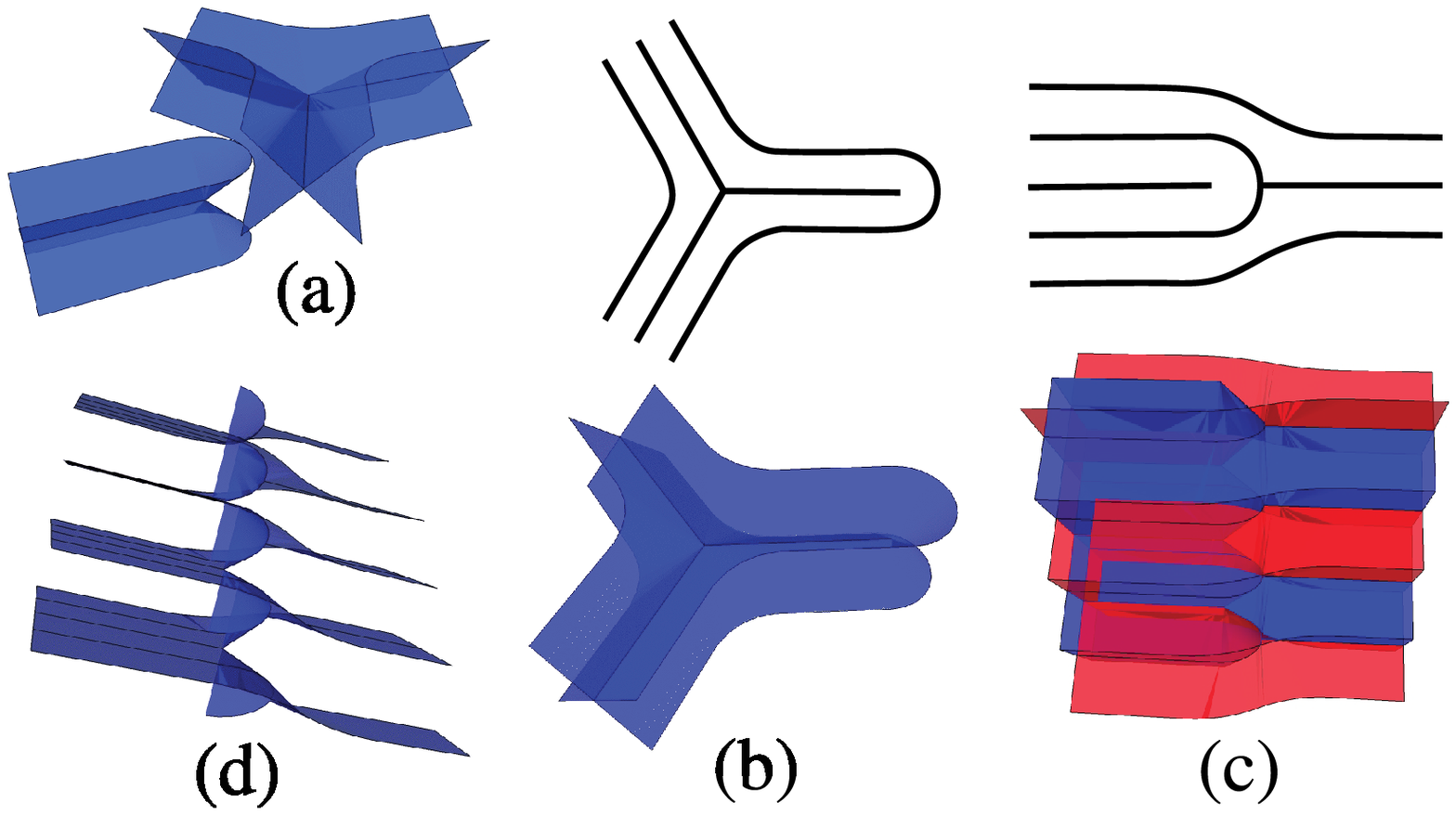}} 
\caption{Disclination dipoles in smectics as examples of more complicated structures. (a) a $+1/2$ and $-1/2$ disclination pair forming a dislocation. For (b) and (c), the top row shows the layers derived from slicing the surface shown in the bottom row at integer values of the height, i.e. taking level sets. (b) pincement, formed from a disclination dipole where the singular points lie at the same height. (c) dislocation, also formed from a disclination dipole, but with the singular points at different heights. (d) a cutaway view of the helicoid sitting inside the dislocation.} 
\label{complicated}
\end{figure*}


\begin{thebibliography}{10}
\bibitem{taylor34}
Taylor GI (1934) The mechanism of plastic deformation of crystals. Part I. Theoretical. {\em Proc R Soc Lond A} 145: 362-387.

\bibitem{kosterlitz73}
Kosterlitz JM, Thouless DJ (1973) Ordering, metastability and phase transition in two-dimensional systems. {\em J Phys C: Solid State Phys} 6: 1181-1203.

\bibitem{toner81} 
Toner JT, Nelson DR (1981) Bond-orientational order, dislocation loops and melting of solids and smectic-A liquid crystals. {\it Phys Rev B} 24: 363-387.

\bibitem{musevic06}
Musevic I, Skarabot M, Tkalec U, Ravnik M, Zumer S (2006) Two-dimensional nematic colloidal crystals self-assembled by topological defects. {\em Science} 313: 954-958.

\bibitem{klemanmicheltoulouse}
Kl\'{e}man M, Michel L, Toulouse G (1977) Classification of topologically stable defects in ordered media. {\em J Phys Lett} 38: L195-L197.

\bibitem{mermin}
Mermin ND (1979) The topological theory of defects in ordered media. {\em Rev Mod Phys} 51: 591-648.

\bibitem{michel}
Michel L (1980) Symmetry defects and broken symmetry. Configurations Hidden Symmetry. {\em Rev Mod Phys} 52: 617-651.

\bibitem{trebin}
Trebin H-R (1982) The topology of non-unifrom media in condensed matter physics. {\em Adv in Phys} 31: 195-254.

\bibitem{poenaru77} 
Po\'{e}naru V, Toulouse G (1977) The crossing of defects in ordered media and the topology of three-media. {\em J Physique} 38: 887-895.

\bibitem{sethna92}
Sethna JP, Huang M (1992) in {\em 1991 Lectures in Complex Systems}
(Eds. Nagel L, Stein D, Santa Fe Institute Studies in the Science of
Complexity, Proc Vol XV, Addison-Wesley), pp~267-288.

\bibitem{chaikinlubensky}
Chaikin PM, Lubensky TC (1995) in {\em Principles of Condensed Matter Physics} (Cambridge University Press, Cambridge), pp~308-315; 495-526.

\bibitem{sethna}
Sethna JP (2008) in {\em Entropy, Order Parameters, and Complexity} (Oxford University Press, Oxford), pp~194-200.

\bibitem{klemanmichel}
Kl\'{e}man M, Michel L (1978) Spontaneous Breaking of Euclidean Invarinace and Classification of Topologically Stable Defects and Configurations of Crystals and Liquid Crystals. {\em Phys Rev Lett} 40: 1387-1390.

\bibitem{kleman}
Kl\'{e}man M, Friedel J (2008) Disclinations, dislocations and continuous defects: A reappraisal. {\em Rev Mod Phys} 80: 61-115.

\bibitem{poenaru}
Po\'{e}naru V (1981) Some aspects of the theory of defects of ordered media and gauge fields related to foliations. {\em Commun Math Phys} 80: 127-136.

\bibitem{nyeberry}
Nye JF, Berry MV (1974) Dislocations in Waves. {\em Proc Roy Soc Lond A} 336: 165-190.

\bibitem{harrison}
Harrison C et al. 
(2002) Dynamics of pattern coarsening in a two-dimensional smectic system. {\em Phys Rev E} 66: 011706.

\bibitem{ruppel}
R\"{uppel} D, Sackmann E (1983) On defects in different phases of two-dimensional lipid bilayers. {\em J Physique} 44: 1025-1034.

\bibitem{kivel} Kivelson SA, Fradkin E, Emery VJ (1998) Electronic liquid-crystal phases of a doped Mott insulator. {\em Nature} 393: 550-553.

\bibitem{frank58} 
Frank FC (1958) On the theory of liquid crystals. {\it Discuss Faraday Soc} 25: 19-28.

\bibitem{kuriklavrentovich}
Kurik M, Lavrentovich OD (1988) Defects in liquid crystals: homotopy theory and experimental studies. {\em Sov Phys Usp} 31: 196-224.

\bibitem{klemanlavrentovich}
Kl\'{e}man M, Lavrentovich OD (2003) in {\em Soft Matter Physics: An Introduction} (Springer-Verlag, New York), pp~447-452.

\bibitem{lowmanohar}
Low I, Manohar AV (2002) Spontaneously broken symmetries and
Goldstone's theorem. {\em Phys Rev Lett} 88: 101602.

\bibitem{gauge_comment}
The astute reader might wonder about the restriction of paths for a spontaneously broken gauge theory where the Goldstone mode is also absent. There are no restrictions: in the gauged case the winding is not a non-trivial loop in a ground state manifold connecting equivalent but distinct ground states, but rather a winding on a gauge orbit connecting different representations of the same physical state. 

\bibitem{comment0}
Here and throughout $\nabla$ acts on the two-dimensional $xy$-plane.

\bibitem{comment}
More precisely we are considering the $n$-jet of the configuration at $P$. Strictly speaking, for $1$-jets, we should keep track of the magnitude $|\nabla\phi|_{{\bf x}=P}$ as well. But since $|\nabla\phi|$ is an element of $\mathbb{R}$, which is contractible, the topological behavior of this component would be trivial. An argument of Kleman, Michel, and Toulouse~\cite{klemanmicheltoulouse} (suitably reinterpreted) shows that jets higher than first order are not needed for a similar reason.

\bibitem{golubitsky}
Golubitsky M, Guillemin V (1973) {\em Stable Mappings and Their
Singularities} (Springer-Verlag, New York).

\bibitem{milnor}
Milnor J (1964) {\em Morse Theory} (Princeton University Press,
Princeton).

\bibitem{farber}
Farber M (2004) {\em Topology of closed one-forms}, (American
Mathematical Society, Providence), in particular see p. 37.

\bibitem{goresky}
Goresky M, MacPherson R (1988) {\em Stratified Morse Theory}
(Springer-Verlag, New York).

\bibitem{berry}
Berry MV (1981) in {\em Physics of Defects}, eds Balian R, Kleman M, Poirier J-P (North-Holland, Amsterdam), pp~453-549. 

\bibitem{comment2}
Note that although the branch point of a helicoid is quite apparent when viewed in three dimensions, in any given slice of it, the branch point is not apparent since the layers all remain smooth, see particularly, images in~\cite{nyeberry}.

\bibitem{fnote}
There may be other types of deformations which we choose to exclude due to energetic or other physical concerns. For instance, it might be convenient to consider deformations up to those which preserve the number of layers, which would prevent us from pushing the surfaces in the vertical direction.

\bibitem{HN}
Halperin BI, Nelson DR (1978) Theory of Two-Dimensional Melting. {\em Phys Rev Lett} 41: 121-124.

\bibitem{riemannref}
Colding TH, Minicozzi WP, II (2006) Shapes of Embedded Minimal Surfaces. {\em Proc Natl Acad Sci USA} 88: 11106-11111.

\bibitem{SVKN}
Santangelo CD, Vitelli V, Kamien RD, Nelson DR (2007) Geometric Theory of Columnar Phases on Curved Substrates. {\em Phys Rev Lett} 99: 017801.

\bibitem{willmore}
Willmore T (1965) Note on Embedded Surfaces. {\em Anal Stunt ale Univ Iasi Sect I a Math} 11: 493-496.

\bibitem{cac}
Nistor AI (2009) Certain Constant Angle Surfaces Constructed on Curves. arxiv:0904.1475v1 [math.DG].

\bibitem{xing}
Xing X (2008) Topology and Geometry of Smectic Order on Compact Curved Substrates. arxiv:0806.2409v2 [cond-mat.soft].

\bibitem{strebel}
Strebel K (1984) {\em Quadratic differentials} (Springer-Verlag, New York).
\end{thebibliography}
\end{document}